# Knowledge Acquisition from Social Platforms Based on Network Distributions Fitting

Jarosław Jankowski[1], Radosław Michalski[2], Piotr Bródka[2], Przemysław Kazienko[2], Sonja Utz[3]

Faculty of Computer Science and Information Technology[1]
West Pomeranian University of Technology
Żołnierska 49, 71-410 Szczecin, Poland
tel. +48 91 449 56 68
fax. +48 91 487 08 42
jjankowski@wi.zut.edu.pl

Institute of Informatics, Wrocław University of Technology[2]
Wrocław, Poland
radoslaw.michalski@pwr.edu.pl, piotr.brodka@pwr. edu.pl, kazienko@pwr.edu.pl

Knowledge Media Research Center[3]
Tübingen, Germany
s.utz@iwm-kmrc.de

**Abstract.** The uniqueness of online social networks makes it possible to implement new methods that increase the quality and effectiveness of research processes. While surveys are one of the most important tools for research, the representativeness of selected online samples is often a challenge and the results are hardly generalizable. An approach based on surveys with representativeness targeted at network measure distributions is proposed and analysed in this paper. Its main goal is to focus not only on sample representativeness in terms of demographic attributes, but also to follow the measures distributions within main network. The approach presented has many application areas related to online research, sampling a network for the evaluation of collaborative learning processes, and candidate selection for training purposes with the ability to distribute information within a social network.

## 1     Introduction

Social networking sites are used as the research environment, and they provide opportunities to analyze real-world behavior (Abbasi et al., 2012) as well as online activities (Utz, Beukeboom, 2011, Gjoka et al., 2009) with the applications in the areas related to collaborative learning (Kwon, Liu, & Johnson, 2014), computer-mediated educational environments (Rummel, Spada, 2005) and knowledge management (Ordóñez de Pablos, 2004). Due to the complexity of the network structures, the analyses are usually performed using some samples to find structures that are smaller, but which share similar properties and distributions (Ebbes et al., 2008). Recent studies in this field have focused on new algorithms (Lee, Kim, & Jeong, 2006, Stumpf, Wiuf, & May, 2005) and various areas of application (Gjoka et al., 2009, Lakhina et al., 2001, Rusmevichientong et al., 2001). The knowledge gathered from social network analysis can be extended using either typical surveys or new approaches based on adaptive surveys that optimize survey costs, quality and response rates. Research in this area is still in the early stages and adaptive methods are rarely implemented (Schouten, Calinescu, & Luiten, 2011). Another motivation for further research on the development of sampling methods is to increase the representativeness of survey data. The majority of studies on social media focuses on social network sites such as Facebook, and many of these studies use (online) surveys (Back, Stopfer, & Vazire, 2010, Utz, Krämer, 2009). The participants are usually students or self-selected. A problem with this approach is the representativeness of the sample - young, highly educated individuals or highly motivated users are usually overrepresented. Similar issues were identified in the field of knowledge management and collaborative learning to build groups with specific profile (Dascalua et al., 2014). Although it is possible to extract behavioral data from social media and use them as the basis of the analysis (Thelwall, 2008, Liu, 2007), social scientists are often interested in the subjective experience of social media users, such as motivations for and gratifications of social media use, evaluation of competences and knowledge resources within the network (Ordóñez de Pablos, 2004, Różewski, Ciszczyk, 2009, Colomo-Palacios et al., 2014b). To evaluate them, surveys are still the most suitable tool. In this paper, a new method for judging and enhancing the representativeness of an online sample is presented. The authors argue that it might be useful to utilize network measures such as centrality or degree as a basis for determining the representativeness of an online sample vs. the entire population.

Some users have a very central social position within the online social networks, and they possess many more inbound and outbound connections when compared with other users. By comparing the network profile of the sample and the overall population, the representativeness of the online sample can be determined. Moreover, it is possible to develop algorithms that suggest which users should be approached in order to enhance the representativeness of a given sample so that the results will have higher potential in the areas of community building, information dissemination, and collaborative learning (Cowan, Jonard, 2004). The approach presented below is based on selecting an adequate set of candidates in each step of the multistage process to improve the representativeness of the sample in terms of network measures. Depending on the research goal and the area of applications, different network characteristics might be considered. To identify opinion leaders, the best candidates for leadership in collaborative learning or knowledge brokers, it is usually necessary to evaluate centrality measures (Boari, Riboldazzi, 2014). However, fulfilling a bridge position is more important when focusing on advertising and diffusing innovation or spreading knowledge among network nodes. From the perspective of collaborative learning, it is important to select nodes with specific characteristic for future activity within the network, and representative selection can impact on the future spread of knowledge within it.

While the structure of connections within the social network influences collaborative learning processes, there is a clear need to access information about participants and their potential for learning processes and sharing of information with other participants. Collaborative learning and group-based learning is closely related to dynamic social systems (Strijbos, 2001) where the members of the community interact and share experiences with one another (Chiu, 2008). During the learning process, members of the community evaluate other ideas and get engaged in monitoring the tasks and progress of other participants (Chiu, 2000). Key problems found here can be addressed to quantify proper users' features, select users with specific characteristic, and split users into optimal groups (Long, Qing-hong, 2014) in order to boost the sharing of knowledge in organizations (Lytras, Tennyson, Ordóñez de Pablos, 2008). During collaborative learning processes, building teams and increasing potential by acquiring additional representatives with specific knowledge or competences can be very important, not only in terms of knowledge itself, but also in terms of network characteristics. While the ability to attain knowledge from all nodes of a network can be limited, sampling methods can be applied to acquire information desired. The proposed method can be adapted to different research goals by using weighted sampling. As online surveys are usually based on voluntary participation, and because there may be low response rates, the obtained sample may have other characteristics than the random sample. The proposed method makes it possible to direct the selection process towards expected characteristics of the sample.

## 2    Related work

*Conventional and adaptive network sampling*

Research related to network sampling is based on various techniques using both conventional and adaptive approaches. Sampling design is treated as *conventional* when it does not use acquired data in the sampling process. The first group of methods in this class is based on random-node selection focused on uniform or proportional-to-node degree probabilities (Maiya, Berger-Wolf, 2010), random edge selection (Ahmed, Neville, & Kompella, 2011) and the egocentric method (Ma et al., 2010). The other group is based on graph sampling and includes snowball sampling (Frank, Snijders, 1994, Frank, 1979) random walk (Thompson, 2006), the forest fire method (Leskovec, Faloutsos, 2006) and others. Apart from theoretical work, some studies were conducted using real online social systems like Facebook (Gjoka et al., 2009) or Twitter (Ahn et al., 2007).

In contrast to static designs, *adaptive* sampling can be applied after the results of earlier stages are collected, and it is used to direct sampling (Handcock, Gile, 2010, Thompson, 2011). Conventional methods have problems with sampling hidden populations, but the adaptive method can change sampling direction on the fly, if necessary. There are approaches targeted to adaptive cluster sampling based on the selection of neighbors in the network only if a given condition related to cluster location is satisfied (Thompson, 1998). Other dedicated methods are used to sample network node selection and the estimation of information diffusion processes in either single-layer (Jankowski, Michalski, & Kazienko, 2012) or multilayer networks (Michalski, Kazienko, & Jankowski, 2013).

The respondent-driven sampling was introduced by Heckathorn (2004) and extended later (Salganik, Heckathorn, 2004). It is based on recruitment of members of the population by other sampled members. Respondent-driven sampling is an extension of snowball sampling and the patterns of recruitment are used to calculate inclusion probabilities for different types of nodes. It collects information about ties from each participant (Heckathorn, 2004), but can be inaccurate in clustered networks because of homophily and separated communities. The proposed adaptive approach is based on the collection of network data from respondents, and adaptive sampling (Thompson, Seber, 1996) is based on moving to other regions of the network after obtaining enough samples from the identified cluster.

*Adaptive approaches to survey design*

While sampling delivers information about the network evolution of data collection methods, new technologies provide possibilities for survey design that were unavailable earlier (Deville, Tillé, 2005). Surveys can be identified as *static* if they are not dependent on collected observations, while *adaptive* surveys are partially based on data from observations (Schouten, Calinescu, & Luiten, 2011). Adaptive surveys are a means of increasing responses and the quality of the research by selecting samples characterized by the lowest mean square error on the sample values. Apart from sampling direction, other adaptive components can be: offering different incentives, using responsive survey designs (Groves, Heeringa, 2006) or questionnaire structures (Singh, Howell, & Rhoads, 1990). Survey adaptation can be based on time intervals between calls, visits and other forms of communication with respondents (Greenberg, Stokes, 1990), survey errors (Lyberg et al., 1997) and survey costs (Groves, 1989). The design-based approach to survey sampling uses variables of interest as fixed values, while model-based variables of interest are defined as random variables with joint distribution (Thompson, 1998). During surveys, interventions can be made to decrease variances of selected variables in the respondent pool by targeting sampling to key subgroups (Couper, Groves, 2009).

Earlier research showed how to optimize the survey process and increase response rates (Schouten, Calinescu, & Luiten, 2011). Schouten et al. systematized adaptive survey designs and provided a mathematical framework to improve the process of data collection based on surveys. Furthermore, the authors defined *Q(p)* as an indicator of quality and *C(p)* as a cost indicator, and optimization was defined as *max*p *Q(p)* with *C(p)* < *C*max and *C*max as maximum budget constraints or *min*p *C(p)* with *Q(p)* >= *Q*min and *Q*min defined as minimum quality. The quality functions can be measured using population data as covariate-based and target survey variables as item-based. The authors considered different strategies with the goal of increasing response rates as a function of the budget. Additionally, the main elements that should be included in adaptive design were identified, including factors affecting costs and error indicators, monitoring of indicators in the initial phase of the survey, altering the survey design during the sampling process and aggregating data from different phases into a single indicator (Couper, Groves, 2009). The advantages of adaptive survey designs are pointed out in several dimensions. Samples built using adaptive approaches can be targeted to desired distributions of attribute values and follow the assumptions of balanced sampling (Tillé, Favre, 2004). Researchers emphasize the ability to increase the performance of surveys by focusing on rare characteristics and targeting hidden populations (Heckathorn, 2004), by the ability to reduce non-responsive bias (Wagner, 2008) and by the reduction of survey costs (Groves, 1989).

*Assumptions for sampling targeted at the distributions of network measures*

The review showed the main areas of the research and applications of adaptive approaches in the field of network sampling and surveys (Lee, Kim, & Jeong, 2006, Leskovec, Faloutsos, 2006). Most of the presented methods that target surveys are focused on sample representativeness in terms of attributes, but the problem that can arise is the representativeness of the surveyed users in terms of network distributions. The majority of studies focus on representativeness of the population in terms of demographic values or other attributes and does not take into account the parameters of the analyzed social networks (Schouten, Calinescu, & Luiten, 2011). The standard methodology does not ensure that all possible knowledge is used while building the representative data sample. Additionally, there is the question of what a representative sample means. Should researchers care only about attributes or about network parameters and their distributions too? Moreover, researchers need to decide which network measures ought to be considered. Finally, the sample built using only attribute distributions may not be representative in terms of network measures and their distributions.

Surveys are usually based on small samples of the network and are volunteer-based. Usually, the whole sample of the acquired data is used for analysis. In such cases, some types of nodes can be overrepresented because the most active users are the main participants in the surveys. This approach raises a possibility: Maybe only the carefully selected part of the acquired data should be utilized and some data should be removed from the final set. Additional feedback from the newly selected nodes needs to be acquired using the adaptive approach.

Other requirements are related to cases in which we know distributions of the several network measures and the sample should keep them. It might be possible to find a "compromise" sample with the same distance to all distributions and the sample should be representative for all available distributions with given weights.

The main goal of this research is to demonstrate a new approach to generating a network sample by selecting nodes for surveying in such a manner that they will follow the required several distributions. The general idea is to minimize the distance between the surveyed sample and the whole network profile in terms of the Kullback-Leibler measure (Kullback, Leibler, 1951). It will also be emphasized that the network sampling approach should result in conformity with network measure distributions, following in many cases the power law, which significantly differs from typical attributes of users with the Gaussian distribution. In a natural way, the available sampling methods are targeted to nodes with a high number of representatives. Specific nodes will not be selected in the samples and some concentration of the samples can be observed.

The method proposed in this paper makes it possible to perform long tail sampling in order to get representatives from areas that are not easily reached by random sampling and other methods. Another assumption is to define a separate weight for each network measure and to force sampling to focus on a selected network measure or set of measures, which makes it possible to obtain a sample specific to the area of application that the researcher needs.

*Areas of application*

One of the applications of the proposed method is the selection of network nodes using characteristics that are important for the diffusion of information within the network; here, features other than node degree play an important role (Boari, Riboldazzi, 2014) and their activity (Jankowski, Michalski, Kazienko, 2013). Information gathering about knowledge within the network can be targeted toward the selection of suitable candidates for collaborative learning as well as those who can increase the potential of an organizational social network in the area of knowledge management (Lytras, Sakkopoulos, & Ordóñez de Pablos, 2009) and for providing knowledge recommendations (Colomo-Palacios et al., 2014a).The proposed approach integrates additional sources of information into the sampling process and can be used within software platforms that are targeted to knowledge gathering and management (Lytras, Ordóñez de Pablos, 2011, Colomo-Palacios et al., 2014c) and in processes of social collaboration and knowledge sharing acceleration (Różewski, 2010).

Another area being addressed is the evaluation of knowledge within a network through the use of testing methods in order to acquire a general view of the current level of knowledge and the potential to spread this knowledge among network members. The other application field is related to selecting proper candidates for supplementary learning with a high probability that they will then act as hubs for spreading knowledge within the network. Training processes within the organization usually require the selection of a relatively small number of workers for additional training. It is possible to select by means of the proposed method the users with the high potential for effective learning according to the given network measures. Additionally, new approaches based on the adaptation of a teaching activity may be developed (Chiu, 2004).

## 3    Conceptual framework

In this part a balanced adaptive distribution fitting approach based on a set of network measure distributions is proposed. Its main goal is to build representative survey responses based on a selected set of participants in terms of distance from the whole network distribution. The function minimizing a distance from the vector of network distributions is proposed, and the network members are selected to fit the reference distributions for the whole network, which are known in advance. In Fig. 1, the sample process of conducting a survey in the social network $SN$ with the proposed approach is presented.

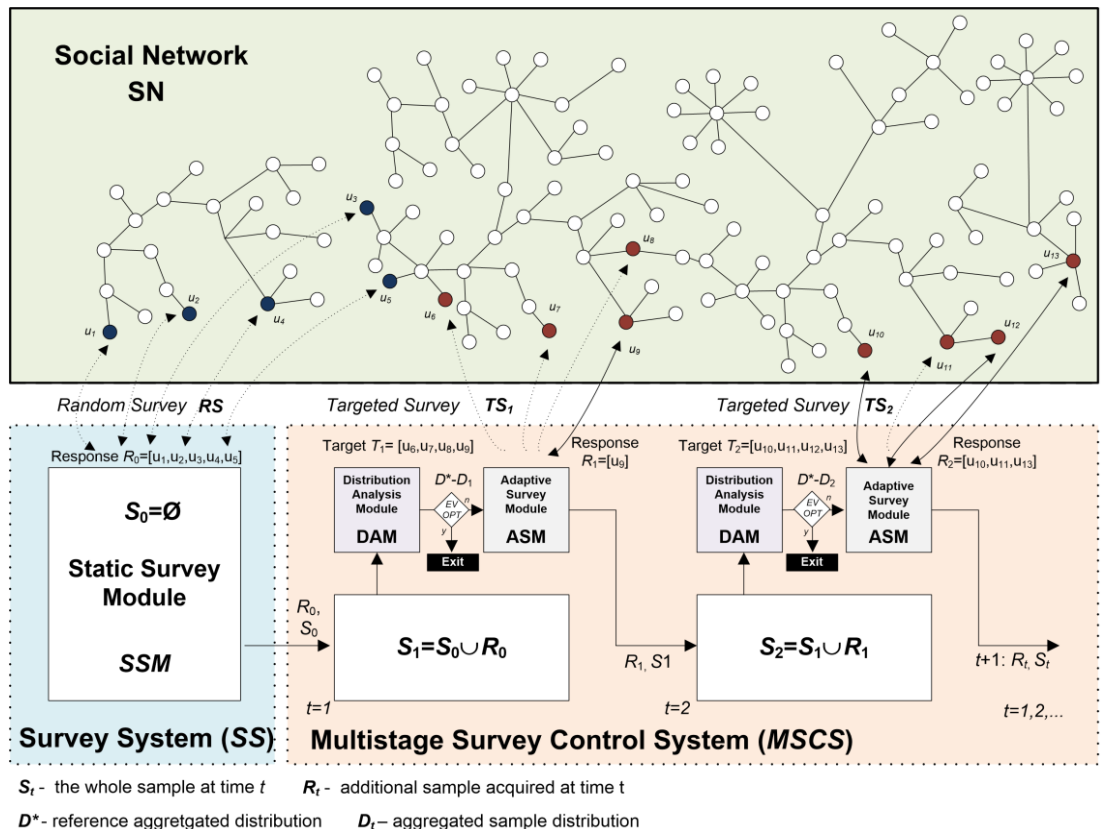

Fig. 1. Sample process controlled by a Multistage Survey Control System

The survey system *SS* with the conventional static approach is responsible for conducting surveys in *SN*. Using the static survey module, *SSM* users fill in a survey on a non-incentivized basis. In the conventional approach, the process ends at this stage and the analyses are carried out on the obtained data sample, even if the sample is not representative. The process illustrated in Fig. 1 has so far obtained data from a set of users $S_0=\{u_1,u_2,u_3,u_4,u_5\}$. Next, the decision is made regarding whether a representative sample should be built from the obtained data by weighting, or if the data collection process should be continued. If it is not possible to collect additional data, the obtained sample is analyzed in terms of network distributions and some selected nodes from the sample with the highest negative impact on overall sample evaluation are excluded from further analyses. The consequence of such an approach would be that only a subset of the obtained data is actually used. If the sampling is continued, a multistage survey control system *MSCS* uses a vector of reference distributions $D^*$ based on the whole social network and a vector of distributions $D_t$ computed for sample $S_t$ at the *t*th stage. If $D_t$ differs from $D^*$ significantly (see Fig. 1), the sample is treated as not representative and additional participants have to be targeted. The distribution analysis module *DAM* and the adaptive survey module *ASM* are included within *MSCS*. The main goal of the *MSCS* is to build a sample $S_t$ that is characterized by small distance from the whole network distribution $D^*$. First, *DAM* is responsible for the distribution analysis of the vector $D_t$. At each stage, the distance from the whole social network distribution is measured by evaluation function $EV_t$ (see Eq. 1). If the value of this function is at an acceptable level ($EV_{opt}$), the process terminates at this stage and the sample $S_t$ can be used for analysis. Otherwise, the DAM system selects the group of candidates in target set $T_t$. These users are requested and those who respond form set $R_t$ which is passed to the next stage.

Fig. 2 depicts how the computations may affect the trajectory of sample *S* to reach the expected level of error – $EV_{opt}$.

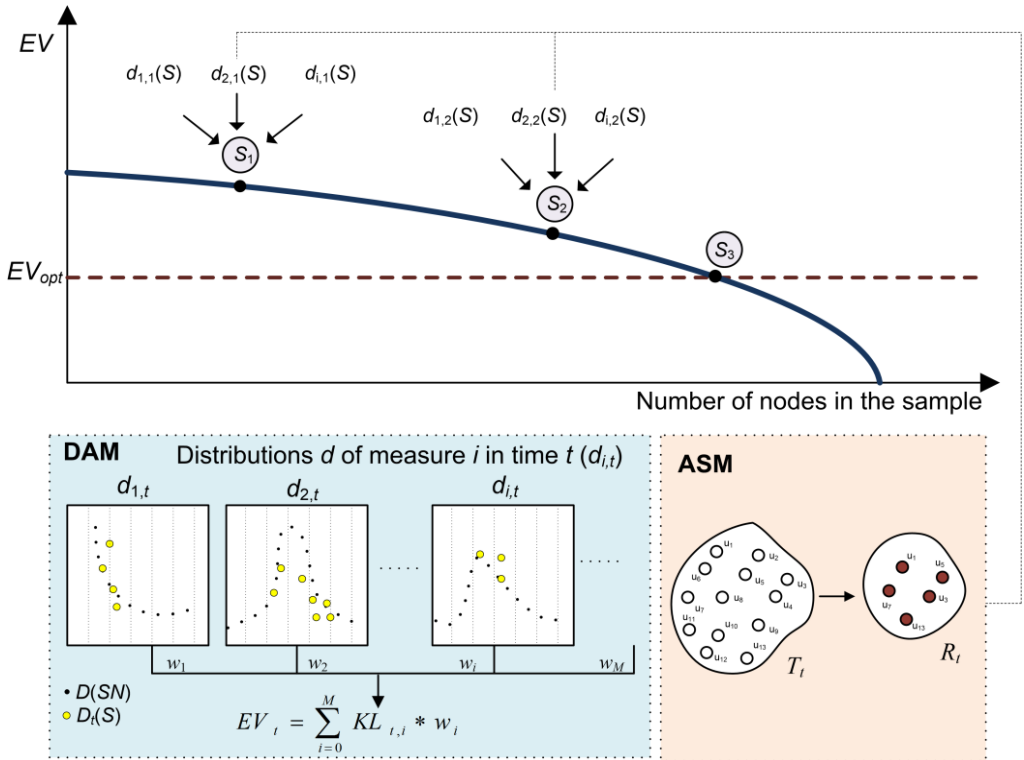

Fig. 2. Adjusting sample with control parameters to fit tunnelling within boundaries

The proposed *MSCS* approach may be represented as a series of the following steps:

1. Stage $t=0$ (static initial survey). For every measure find the best distribution function approximation for the whole network. The resulting vector will consist of functions and their parameters describing the particular reference and expected measure distribution – $D^*$. Next, request the initiate set of target users $T_0$ (selected in any way) and obtain the initial set of those who responded and were surveyed $R_0$, $S_0=\emptyset$.
2. Initiate stage $t$ (validation and extension). Update sample set $S_t$ composed of the users surveyed so far ($S_t=S_{t-1}\cup R_{t-1}$) and calculate the approximation functions as above – $D_t$.
3. Evaluate the difference, i.e. Kullback-Leibler divergence $KL_{t,i}$[42] between two functions: measure distributions from the tested ($D_t$) and reference set ($D^*$), and separately for each network measure $i$ out of all $M$ measures considered.
4. To generalize and evaluate the result globally, calculate the function $EV_t$ in stage $t$, as follows:

$$EV_t = \sum_{i=0}^{M} KL_{t,i} * w_i, \qquad (1)$$

   where $KL_{t,i}$ is the result of recent Kullback-Leibler divergence for the *i*th measure, $w_i$ is a weight representing importance of the *i*th measure, and $M$ is the total number of evaluated measures.
5. If $EV_t > EV_{opt}$ (the goal has not been obtained yet), collect an additional set $R_t$ of responded users (for expansion of $S_{t+1}$) by: (i) selection of the new target $T_t$ using, e.g., the proposed *K*-bins algorithm (see below), where the number of nodes $|T_t|$ is the parameter set by the researcher depending, e.g., on the time allowed $t$, (ii) requesting users from $T_t$ – responding ones form $R_t$; repeat steps 2-5 until $EV_t <= EV_{opt}$.

At the beginning some preliminary decisions shall be made. First, the set of *M* network measures should be defined. In order to calculate network metrics within a reasonable time, it is suggested to focus on local measures (e.g. degree), rather than on global ones (e.g. between-ness). Because various measures may have various distributions, i.e. normal, power law, etc. they cannot be combined. In such case, every measure distribution function in $D_t$ shall be compared to $D^*$ and the comparison result may be combined later. For comparison of distribution functions, the Kullback-Leibler divergence was selected, because it is able to compare different types of distributions. Another issue are possibly different domains of various measures. To be able to make $KL_{t,i}$ values comparable, all the measures should be normalized into a similar domain, e.g. range [0,1]. As described above, the adaptive nature of the algorithm lies in the way of choosing nodes, if the value of $EV_t$ is not satisfying. The authors propose an algorithm developed to reflect the nature of typical social network measure distributions. Its properties are described below.

Algorithm *K*-bins: Selects set $T_t$ of nodes to be surveyed. Introduce the normalized measure *NM*, which combines all *M* evaluated social network measures into a single one by adding their normalized values. Calculate *NM* for all nodes in the network and create its histogram by splitting its domain into *K* bins (adjoining ranges), where *K* is a fixed parameter. For every bin select $|T_t|/N$ percent of nodes in that bin (i.e. $(|T_t|*N_k)/N$ of nodes), equally distributed over the bin, where *N* is the total number of nodes in the network, and $N_k$ is the number of nodes in

the *k*th bin. If the number of nodes in a bin selected for evaluation is less than one, choose one node. Then evaluate the resulting set of representatives.

The proposed *K*-bins algorithm uses equal random sampling in the selected regions of the distribution (bins) to provide better sample representation in terms of the aggregated network measures. Splitting the search space into bins makes it possible to perform sampling in the underrepresented regions, especially if the response rate is low. New nodes can be searched for in selected bins with higher performance rather than sampling the whole network. Typical random sampling requires the entire sample to cover the whole region, while the proposed method can work on smaller samples from areas of distribution where the representation should be improved.

The proposed method is based on the fact that most of the typical social network measures are following the power law. In such a case, by random sampling we would obtain some candidates from the so-called *long tail* of the measure distribution, where it is typical to have a small number of nodes with rare values of network measures. By obtaining some more nodes to be surveyed and using *K*-bins for their selection, one can gather information about rare nodes in the initial steps of the *MSCS* algorithm, as shown in diagram A in Fig. 3, whereas raw simple random sampling may or may not produce any node from *the long tail*.

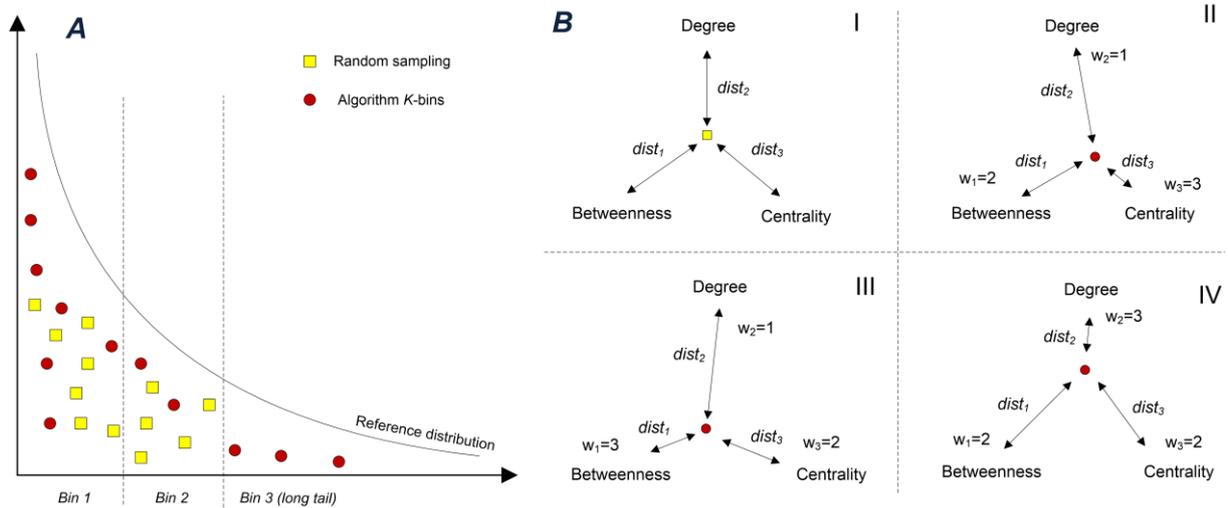

Fig. 3. The proposed method vs. random sampling in terms of distance
from distributions and long tail sampling

The presented approach works similarly to multi-objective optimization methods, where an optimal solution must fulfil the assumed criteria and deliver a solution maximizing or minimizing a set of objective functions. For each objective function, the weight representing desired preferences can be assigned if the same weights are positioned in the central point of the decision space with the same distance to all criteria. By using weights, this point can be moved in this space and the Pareto optimal point can be found when one of the criteria cannot be improved without decreasing the values for the others. In our method, a similar approach is used for sampling. By changing weights it is possible to adjust the sampling process when a higher priority of one of the measures is expected. Advantages of the proposed approach are illustrated in Fig. 3 section B. Random sampling (I) shows equal distance of the sample from all network distributions and approximately $dist_1 = dist_2 = dist_3$, and no change in preferences is possible. For the proposed methods in diagrams II, III and IV, the distances of the sample from distributions can be adjusted using appropriate weights in the algorithm (see Eq. 1).

# 4 Empirical research

The new approach is demonstrated by presenting the results of a survey performed within an online social network based on a graphical virtual world with both entertainment and educational purposes. An online survey covering motivations, self-disclosure and self-presentation was conducted among portal users and filled in by 373 of them, while 9,631 users logged into the system in the examined period and were identified by their unique *user_ID*. The structural measures computed for the full network and for the surveyed users showed significant differences between the whole and the surveyed network profile. See Table 1.

Table 1: Main network measures for full network and surveyed users

|  | Full network | Surveyed users |
| --- | --- | --- |
| Number of nodes | 9631 | 304 |
| Number of edges | 294935 | - |
| Fraction of reciprocal links | 74.97% | - |
| **Average node in-degree** | **30.62** | **99.26** |
| **Average node out-degree** | **30.62** | **89.02** |
| Average node degree | 61.25 | 188.27 |
| Average clustering coefficient | 0.07 | 0.08 |

A preliminary analysis of the data reveals that surveys based on voluntary participation did not necessarily follow random sampling. The obtained dataset would be closer to the random sample with the increasing response rate value. However, due to the low response rate the obtained sample has a profile other than random and it can be additionally improved using the proposed method.

In the next step, this dataset was used to empirically test the approach. Authors analyzed the results from this survey in terms of survey participant representativeness compared to the whole network measure distribution. The results of the survey itself were not analyzed.

*Adaptive sampling results*

The proposed *MSCS* method was evaluated by means of a new algorithm and independently with random sampling as a baseline. The following network structural measures were used: indegree, outdegree, total degree centrality and clustering coefficient. In the first stage of this research, the performance of the proposed algorithm was analyzed for a relatively small fraction of the whole network, i.e. less than 5% of the nodes. The study was based on the evaluation function $EV_t$ representing the distance from the expected distribution (for the whole network) and the recent one (for the surveyed sample). The results showed that for this range, the *MSCS* method with the *K*-bins algorithm significantly outperformed random sampling. See Fig. 4.

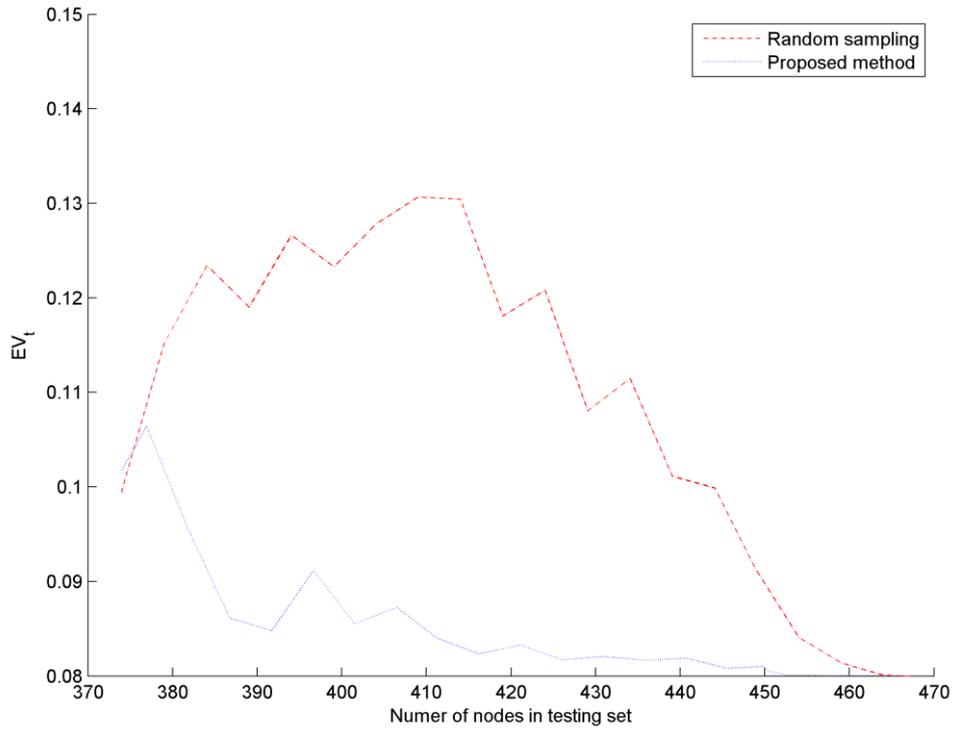

Fig. 4. Random sampling vs. proposed sampling algorithm for measures: indegree, outdegree, total degree centrality for samples less than 5% of all network nodes

Further results presented in Fig. 5 revealed that for samples with a size greater than 5%, the proposed method provides worse results than the random method in terms of $EV_t$. As previously described, the random method with the highest probability acquires the most common nodes, but it may have no knowledge about the end of the curve of power-law distribution until less common nodes are obtained. In contrast, the proposed algorithm tries to obtain this knowledge at the beginning, but at the cost of initial errors, because the algorithm is forced to acquire less common nodes to create a sample representing all the nodes in the network.

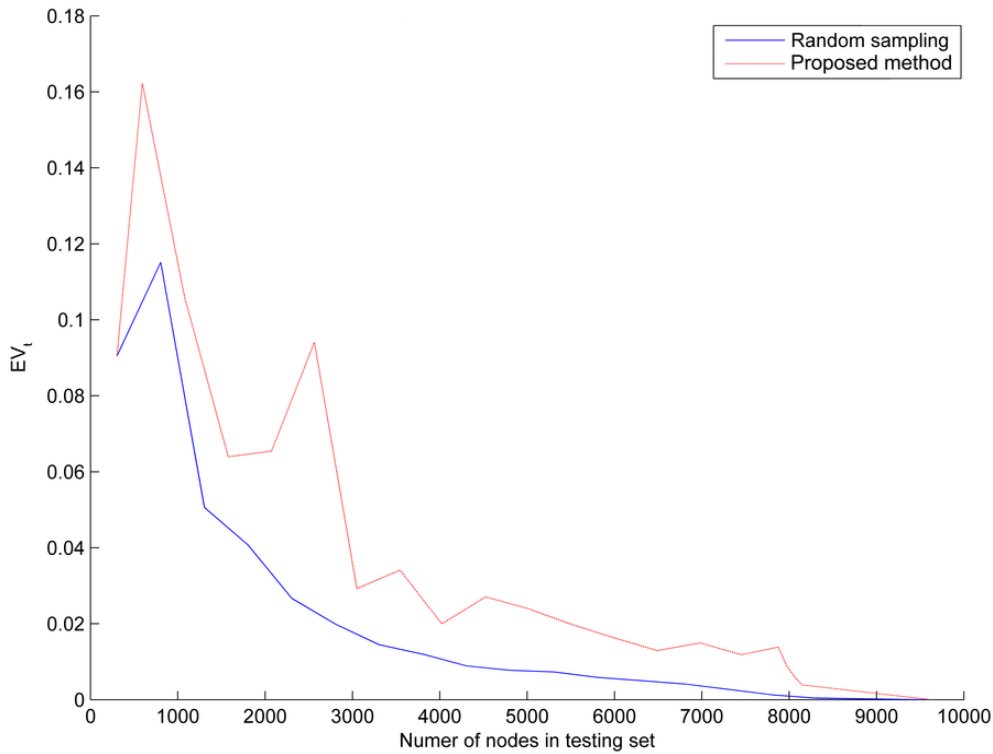

Fig. 5. Random sampling vs. proposed sampling algorithm for measures: indegree, outdegree, total degree centrality for the whole range of samples

It should be noted that in the real world the response rate can be quite low and not all selected nodes will replay. Thus, it may happen that even if a node from the long tail is selected, either by random sampling or by *K*-bins algorithm, it may ignore the request. In such a case our algorithm will just select another node from the long tail in the next iteration, while in the random sampling we might have to wait many iterations to get the next node from the long tail.

*Power law decrease of error*

With the increase of the $S_t$ quantity, the $EV_t$ value does not decrease linearly. This leads to the conclusion that increasing the number of surveyed users is effective only to a point, after which there will be little benefit from gaining additional surveys. The results presented in Fig. 6 refer to random samples. After sampling 10% of the network, $EV_t$ was reduced to 30% and sampling 20% of the network provides a 72% reduction.

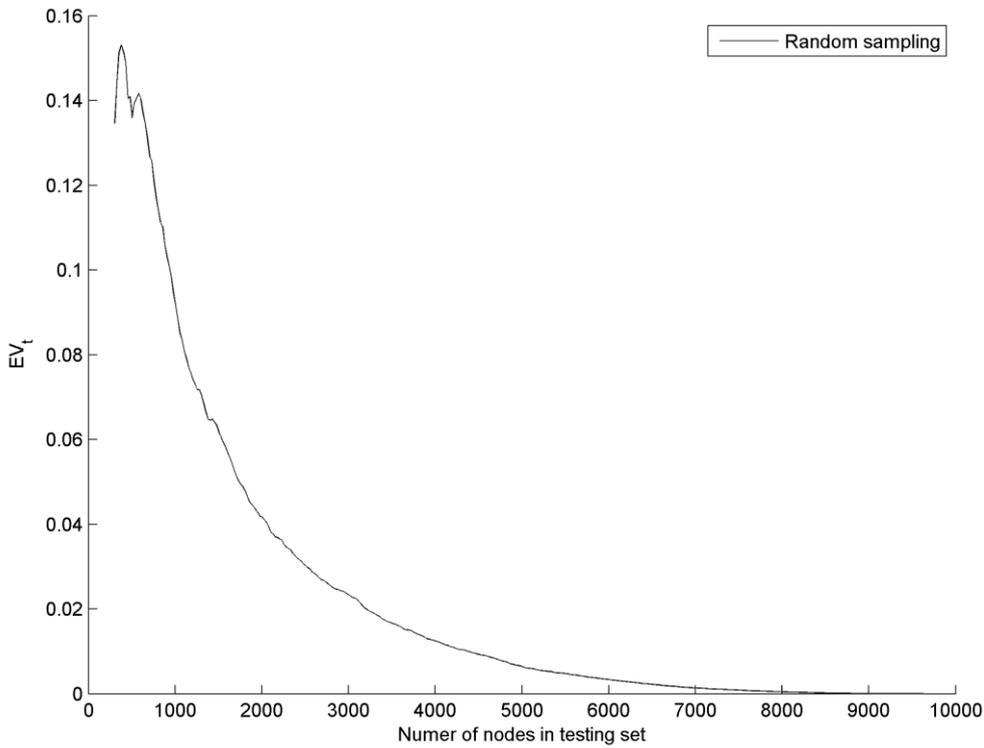

Fig. 6. $EV_t$ values for the tested set starting from the initially surveyed users to the whole network – random sampling – indegree and total degree measures combined

## 5  Discussion and summary

Growing engagement in social network systems and moving from traditional environments to online systems creates a new space for both theoretical and empirical studies. Together with technological development, the need for new methods also grows, making research processes more efficient and increasing their quality. While adaptive survey methodologies were the subject of earlier research, they are not frequently applied to online research. An alternative to available solutions was presented in this paper as an iterative, adaptive getting some more people to be surveyed and it can be implemented within almost any surveying system.

Three algorithms for the uniform sampling nodes were tested during research. The first algorithm aggregated all evaluated measures into a single one by summing all values of those measures. The obtained values are sorted in ascending order and the additional nodes are selected starting from the node with the lowest aggregated measure value and continuing towards the one with the highest value. In the second algorithm, for each element a chosen measure values were calculated and the nodes with its lowest values were chosen. This algorithm was applied in four variants, with different base structural measures: indegree centrality, outdegree centrality, node degree and clustering coefficient. The third algorithm introduced a normalized measure *NM*, which combined all evaluated measures into a single one by summing all normalized values of component values. The value of *NM* was calculated for all the nodes in the network and a histogram representing them with twenty bins was created. If the number of total nodes in the network is *n* and the number of currently surveyed users is *i*, *i*/*n* representatives were chosen for every bin, starting from the borders of the bin and turning towards the middle. The first two algorithms were found to over-represent the users most often found in the network, because most local metrics follow the power law distribution. The third proposed algorithm modeled the referencing distribution adequately and was presented in paper in details.

The results of the experiments lead to interesting conclusions. They have shown that random sampling does not always deliver a sample adequate to the network measure distributions, especially in the case of multiple combined measures and if the sample is relatively small (up to 5% of nodes). Another observation is that after reaching a specified number of users, the effectiveness of further sampling decreases drastically. Based on knowledge of the entire population and by introducing the proposed *MSCS* method and the *K*-bins algorithm, it is possible to get results faster and with less distance from the compound network distributions than by using typical random sampling. The main limitation of the proposed approach is the availability of measure distributions for the whole network, so the method can easily be implemented by online social network operators or others with knowledge about distributions. Note that even if an online social network operator is not able to extract representative random samples, they could know the network distributions and may apply the *MSCS* method.

In a typical social network system, researchers run surveys online, gather results and analyze them, usually with the assumption that the results are valid for the whole social network. The presented *MSCS* approach uses the full network distribution as a reference. It facilitates implementing a dynamic value of incentives in the following stages to obtain more valuable, i.e. representative, samples for research. If more surveyed nodes are expected, higher incentives can be introduced and the value of incentives can be adjusted based on the observed responses.

The proposed approach has applications in several areas. For marketing purposes, the network measures and their distributions can be important along with the attributes of the nodes. If surveys are performed within an organization and are targeted at knowledge management or collaborative learning, the sampling process can be directed toward selecting candidates with a high likelihood of increasing their own competences and sharing knowledge with others in the network. Therefore, the proposed approach can be used while selecting candidates for training. If selecting nodes based on learning potential, network measures should also be taken into account, as selecting only high-degree nodes may result in subjects with lower scores in terms of betweenness or closeness. The proposed approach can help an organization be more cost effective by, for example, using its available funds on only the 5% of workers who have been recommended for training. This makes it possible to select a set of nodes with high potential for spreading information within a given network; which can increase the effectiveness of various processes. In collaborative learning processes, several social network parameters have a high influence on their effectiveness. Individuals with high position in the network can take an important role in these processes. It is not only related to node degree, but also to other network measures. Independently, the proposed approach can be used for the selection of network members to develop teams with specific profile for collaborative learning. It can be used to identify network learning or teaching abilities using samples drawn from a bigger network. It can be important not only to have unrepresented users with high degree, like what is found in typical sampling, but also to have users with other potentially beneficial qualities.

Together with technological development, there is a growing need for new methods that increase the efficiency and quality of research processes. *MSCS* is a promising technique to increase the representativeness of online samples while at the same time reducing effort and costs by targeting the most representative users for areas of applications listed above. While this paper showed a new approach in the methodological level, i.e. iterative getting new respondents independently for distribution regions of network measures, some other algorithms and methods can be created to follow this idea. Future research includes the utilization of node attributes and combining them with evaluations based on network measurements. The proposed algorithms may still require further tuning to address the question of how to design better iterative sampling algorithms leading to faster error reduction, e.g. by using non-equal sizes of bins or various measures of distance.

**Acknowledgments**

The work was partially supported by Fellowship co-Financed by European Union within European Social Fund, by European Union's Seventh Framework Programme for research, technological development and demonstration under grant agreement no 316097 [ENGINE] and by The National Science Centre, the decision no. DEC-2013/09/B/ST6/02317.

**References**

Abbasi M. A., Chai S. K., Liu H., Sagoo K., (2012). Real-World Behavior Analysis through a Social Media Lens, Social Computing, Behavioral - Cultural Modeling and Prediction. Lecture Notes in Computer Science, vol. 7227, pp. 18-26.
Ahmed N., Neville J., Kompella R., (2011). Network Sampling via Edge-based Node Selection with Graph Induction, Computer Science Technical Reports, Purdue University, Technical Reports. TR-11-016.
Ahn Y. Y., Han S., Kwak H., Eom Y. H., Moon S., Jeong H. (2007). Analysis of Topological Characteristics of Huge Online Social Networking Services. In: Proceedings of the 16th international conference on World Wide Web. ACM Press, New York, pp. 835-844.
Back M., Stopfer J. M., Vazire S., (2010). Facebook profiles reflect actual personality, not self-idealization. Psychological Science, vol. 21(3), pp. 372-374.
Boari C., Riboldazzi F., (2014). How knowledge brokers emerge and evolve: The role of actors' behaviour. Research Policy, vol. 43(4), pp. 683-695.
Chiu M. M., (2004). Adapting teacher interventions to student needs during cooperative learning. American Educational Research Journal, vol. 41, pp. 365-399.
Chiu, M. M. (2000). Group problem solving processes: Social interactions and individual actions. Theory of Social Behavior, vol. 30(1), pp. 27-50.
Chiu, M. M. (2008). Flowing toward correct contributions during groups' mathematics problem solving: A statistical discourse analysis. Journal of the Learning Sciences, vol. 17(3), pp. 415 - 463.
Colomo-Palacios R., Casado-Lumbreras C., Soto-Acosta P., Misra S. (2014a). Providing knowledge recommendations: an approach for informal electronic mentoring. Interactive Learning Environments, vol. 22(2), pp. 221-240.
Colomo-Palacios R., González-Carrasco I., López-Cuadrado J. L., Trigo Ribeiro A., Varajao, J. (2014b). I-COMPETERE: Using Applied Intelligence in search of competency gaps in software project managers. Information Systems Frontiers, vol. 16(4), pp. 607-625.
Colomo-Palacios R., López-Cuadrado J. L., González-Carrasco I., García-Peñalvo F. J., (2014c). SABUMO-dTest: Design and Evaluation of an Intelligent collaborative distributed testing framework. Computer Science and Information Systems, vol. 11(11), pp. 29-45.
Couper M., Groves R., (2009). Moving from prespecified to adaptive survey design. Presented at the Modernization of Statistics Production in Stockholm, Sweden.
Cowan R., Jonard N., (2004). Network structure and the diffusion of knowledge. Journal of Economic Dynamics and Control, vol. 28(8), pp. 1557-1575.

Dascalua M. I. , Bodea C. N. Lytras M., Ordoñez de Pablos P., Burlacua A., (2014). Improving e-learning communities through optimal composition of multidisciplinary learning groups, Computers in Human Behaviour, vol. 30, pp. 362-371.

Deville J. C. , Tillé Y., (2005). Variance approximation under balanced sampling. Journal of Statistical Planning and Inference, vol. 2, pp. 569-591.

Ebbes P., Huang Z., Rangaswamy A., Thadakamalla H.P., (2008). Sampling Large-scale Social Networks: Insights from Simulated Networks. In: Proceedings of the 18th Annual Workshop on Information Technologies and Systems, Paris, France.

Frank O., (1979). Estimation of population totals by use of snowball samples. In: Holland P, Leinhards S, eds. Perspectives on Social Network Research. New York, Academic Press, pp. 319-347.

Frank O., Snijders T., (1994). Estimating the size of hidden populations using snowball sampling. Journal of Official Statistics, vol. 10, pp. 53-67.

Gjoka M., Kurant M., Butts C.T., Markopoulou A., (2009). Practical Recommendations on Crawling Online Social Networks. IEEE Journal on Selected Areas in Communications (JSAC), Special Issue on Measurement of Internet Topologies, vol. 29(9).

Greenberg B. S., Stokes S. L., (1990). Developing an optimal call scheduling strategy for a telephone survey. Journal of Official Statistic, vol. 6, pp. 421-435.

Groves R. M. , Heeringa S. G., (2006). Responsive design for household surveys: tools for actively controlling survey errors and costs. Journal of the Royal Statistical Society, Series A, vol. 169, pp. 439-457.

Groves R.M., (1989). Survey errors and survey costs. New York, Wiley.

Handcock M. S., Gile K. J., (2010). Modeling social networks from sampled data. Annals of Applied Statistics, vol. 4(1), pp. 5-25.

Heckathorn D. D., (2007). Extensions Of Respondent-Driven Sampling: Analyzing Continuous Variables And Controlling For Differential Recruitment. Sociological Methodology, vol. 37(1), pp. 151-207.

Jankowski J., Michalski R., Kazienko P., (2013). Compensatory Seeding in Networks with Varying Availability of Nodes. The 2013 IEEE/ACM International Conference on Advances in Social Networks Analysis and Mining, IEEE Computer Society, pp. 1242-1249.

Jankowski J., Michalski R., Kazienko P. (2012). The Multidimensional Study of Viral Campaigns as Branching Processes. SocInfo 2012 - The 4th International Conference on Social Informatics. Lecture Notes in Computer Science LNCS, Springer, Berlin Heidelberg, vol. 7710, pp. 462-474.

Kullback S., Leibler R. A., (1951). On Information and Sufficiency. Annals of Mathematical Statistics, vol. 22, pp. 79-86.

Kwon K., Liu Y. H., Johnson L. P. (2014). Group regulation and social-emotional interactions observed in computer supported collaborative learning: Comparison between good vs. poor collaborators. Computers & Education, vol. 78, pp. 185-200.

Lakhina. A, Byers J. W., Crovella M., Xie P., (2003). Sampling biases in IP topology measurements. In: 22nd Annual Joint Conference of the IEEE Computer and Communications Societies, IEEE Press, pp. 332-341.

Lee S. H. , Kim P. J. , Jeong H., (2006). Statistical properties of sampled networks. Physics Review Letters E 2006, vol. 73:016102.

Leskovec J., Faloutsos C., (2006). Sampling from large graphs. In: 12th ACM SIGKDD International Conference on Knowledge discovery and data mining, ACM Press, pp. 631-636.

Liu H., (2007). Social network profiles as taste performances. Journal of Computer-Mediated Communication, vol. 13(1), pp. 252-275.

Long C., Qing-hong Y., (2014). A group division method based on collaborative learning elements. In:Proceedings of The 26th Chinese Control and Decision Conference, pp. 1701-1705.

Lyberg L. E., Biemer P., Collins M., Leeuw E. D., Dippo C., Schwarz N., Trewin D., (1997). Survey measurement and process quality, Wiley Series in Probability and Statistics, New Jersey: Wiley.

Lytras M., Tennyson R., Ordóñez de Pablos P., (2008). Knowledge and Networks: The social software perspective, IGI-Global. Pp 1-422

Lytras M.D., Ordóñez de Pablos, P. (2011), "Software Technologies in Knowledge Society", Journal of Universal Computer Science, vol. 17(9), pp. 1219-1221.

Lytras M.D., Sakkopoulos E., Ordóñez de Pablos P. (2009). "Semantic Web and Knowledge Management for the health domain: state of the art and challenges for the Seventh Framework Programme (FP7) of the European Union (2007-2013)", International Journal of Technology Management, vol. 47, Nos. 1/2/3, pp. 239-249.

Ma H., Gustafson S., Moitra A., Bracewell D., (2010). Ego-Centric network sampling in viral marketing applications. In: Ting IH, Wu HJ., Ho TH, eds. Mining and Analyzing Social Networks. Studies in Computational Intelligence, Berlin, Springer, pp. 35-51.

Maiya A. S., Berger-Wolf T.Y., (2010). Online Sampling of High Centrality Individuals in Social Networks, Lecture Notes in Computer Science, vol. 6118, pp. 91-98.

Michalski R., Kazienko P., Jankowski J., (2013). Convince a Dozen More and Succeed - The Influence in Multi-layered Social Networks. The Second Workshop on Complex Networks and their Applications at SITIS 2013 - The 9th International Conference on Signal Image Technology & Internet based Systems, December 2-5 2013, Kyoto, Japan, IEEE Computer Society, pp. 499-505.

Ordóñez de Pablos P. (2004). "The nurture of knowledge-based resources through the design of an architecture of human resource management systems: implications for strategic management", International Journal of Technology Management, Special Issue, vol. 27(6/7), pp. 533-543.

Różewski P., (2010). A Method of Social Collaboration and Knowledge Sharing Acceleration for e-Learning System: The Distance Learning Network Scenario. In: 4th International Conference on Knowledge Science, Engineering and Management (KSEM), pp. 148-159.

Różewski P., Ciszczyk M. (2009). Model of a collaboration environment for knowledge management in competence based learning. In: 1st International Conference on Computational Collective Intelligence, Semantic Web, Social Networks and Multiagent Systems, pp. 333-344.

Rummel N., Spada H., (2005). Learning to collaborate: An instructional approach to promoting collaborative problem solving in computer-mediated settings. Journal of the Learning Sciences, vol. 14(2), pp. 201–241.

Rusmevichientong P., Pennock D., Lawrence S., Giles C. L., (2001). Methods for sampling pages uniformly from the WWW. In: Proceedings of AAAI Fall Symposium on Using Uncertainty Within Computation, pp. 121-128.

Salganik M. J., Heckathorn D. D., (2004). Sampling and estimation in hidden populations using respondent-driven sampling. Sociological Methodology 2004, vol. 34, pp. 193-239.

Schouten B., Calinescu M., Luiten A. (2011). Optimizing quality of response through adaptive survey designs, Statistics Netherlands 2011, vol. 11, pp. 1-27.

Singh S., Howell R. D., Rhoads G. K. (1990). Adaptive Designs for Likert-Type Data: An Approach for Implementing Marketing Surveys. Journal of Marketing Research 1990, vol. 27(3), pp. 304-321

Strijbos J. W., (2001). Group-based learning: Dynamic interaction in groups. In P. Dillenbourg, A. Eurelings, & K. Hakkarainen (Eds.), Proceedings of the 1st European Conference on Computer-Supported Collaborative Learning, Maastricht, The Netherlands: Maastricht McLuhan Institute. pp. 569-576.

Stumpf M. P. H., Wiuf C., May R. M., (2005). Subnets of scale-free networks are not scale free: Sampling properties of networks. Proceedings of National Academy of Science, vol. 102(12), pp. 4221-4224.

Thelwall M., (2008). Social networks, gender, and friending: An analysis of MySpace member profiles. Journal of the American Society for Information Science and Technology, vol. 59(8), pp. 1321-1330.

Thompson S. K., (2011). Adaptive network and spatial sampling, Survey Methodology, vol. 37(2), pp. 183-196.

Thompson S. K. , Seber G. A. F., (1996). Adaptive Sampling. New York: John Wiley & Sons.

Thompson S. K., (1998). Adaptive sampling in graphs, In: Proceedings of the Survey Research Methods Section, American Statistical Association, pp. 13-22.

Tillé Y., Favre A. C., (2004) Co-ordination, combination and extension of optimal balanced samples. Biometrika, vol. 91, pp. 913-927.

Utz S., Beukeboom C. J., (2011). The Role of Social Network Sites in Romantic Relationships: Effects on Jealousy and Relationship Happiness. Journal of Computer-Mediated Communication, vol. 16(4), pp. 511-527.

Utz S, Krämer N., (2009).  The privacy paradox on social network sites revisited: The role of individual characteristics and group norms. Cyberpsychology: Journal of Psychosocial Research on Cyberspace 2009; vol. 3(2).

Wagner J., (2008). Adaptive Survey Design to Reduce Nonresponse Bias. A doctoral dissertation, University of Michigan.